# Level Up Your Strategy:
# Towards a Descriptive Framework for Meaningful Enterprise Gamification

Umar Ruhi

> *" Good design is making something intelligible and memorable. Great design is making something memorable and meaningful. "*
>
> Dieter Rams
> Industrial Designer

Gamification initiatives are currently top-of-mind for many organizations seeking to engage their employees in creative ways, improve their productivity, and drive positive behavioural outcomes in their workforce – ultimately leading to positive business outcomes on the whole. Despite its touted benefits, little empirical research has been done to date to investigate technological and individual personal factors that determine the success or failure of enterprise gamification initiatives. In this article, we provide a summary of our preliminary research findings from three case studies of gamification initiatives across different business contexts and present an empirically validated descriptive framework that details the key success factors for enterprise gamification. Our adaptation of the mechanics, dynamics, and aesthetics (MDA) framework for enterprise gamification aims to explicate the connections between end-user motivations, interactive gameplay elements, and technology features and functions that constitute effective gamification interventions in the enterprise. Following a discussion of the core elements in the framework and their interrelationships, the implications of our research are presented in the form of guidelines for the management and design of gamification initiatives and applications. The research findings presented in this article can potentially aid in the development of game mechanics that translate into positive user experiences and foster higher levels of employee engagement. Additionally, our research findings provide insights on key success factors for the effective adoption and institutionalization of enterprise gamification initiatives in organizations, and subsequently help them enhance the performance of their employees and drive positive business outcomes.

## Introduction

As a relatively new breed of technology-based intervention, gamification refers to the process of utilizing a digital platform to incorporate game-like elements in non-game contexts with the aim to positively influence user motivation and to improve user engagement in desired behaviours. In an enterprise setting, gamification techniques may be applied to engage employees in helping an organization realize business process improvements, service efficiencies, talent development, innovative research ideas, and constructive collaboration practices.

Although the hype surrounding enterprise gamification has not yet receded, some early adopters have reported failures with gamification initiatives (Burke, 2014). Their experience has afforded more credence to those who question the potential of gamification – whether it constitutes a trivialization of work and whether it is a frivolous diversion.

To counter these negative accounts, analysts and experts have directed attention to the myriad of success stories that demonstrate the benefits of gamification to organizations in various sectors including airlines, healthcare, financial services, consumer products, and

# Towards a Descriptive Framework for Meaningful Enterprise Gamification
*Umar Ruhi*

education (Buggie, 2014; Palmer et al., 2012; Wang, 2011). Consequently, these experts have expounded that organizations and their leaders need to avoid jumping on the gamification bandwagon and not use it in a knee-jerk fashion to coerce behaviour and outcomes. Rather, organizations and leaders are urged to understand the business case for gamification, appreciate the opportunities and limitations associated with it, and approach the implementation of technologies within the firm's specific organizational and individual context. Attention has been drawn to factors – such as business objectives, employee motivations, and user experience – that constitute key determinants in the effective adoption of enterprise gamification programs. However, owing to the novel nature of gamification and its emergent corporate use cases, there is a general dearth of academic and industry literature explaining these issues (Deterding et al., 2013; Hamari et al., 2014).

In this article, we address this research gap by reporting some emergent findings from our ongoing research on enterprise gamification. We investigated gamification initiatives at three case study organizations from different industries, and conducted interviews with strategy and design teams, evaluated the implementation of gamification applications, and surveyed end users from the organizations. Figure 1 summarizes the case study organizations that we surveyed for our research and the specific methods that we followed to obtain data and derive insights about gamification initiatives in these organizations. To preserve confidentiality of information, we only report the general industry of case study organizations using the North American Industry Classification System (NAICS) and provide a generic context of the gamification applications being used by the case study organization.

In the sections that follow, we provide a summary of the preliminary findings from our research program. First, we offer a working definition of meaningful enterprise gamification and summarize its conceptual underpinnings. Next, we discuss our adaptation of the mechanics, dynamics, and aesthetics (MDA) framework – a descriptive framework that highlights various elements of meaningful enterprise gamification, and provides an overall synopsis of strategy, design, and user experience elements from gamification initiatives and applications across the three case study organizations that we surveyed. The framework is geared towards explaining how gamification leverages human psychology using technology platforms and motivates individual behaviours that drive organizational outcomes. Finally, drawing upon the descriptive framework, we provide guidelines for the management of gamification initiatives and the design of gamification applications.

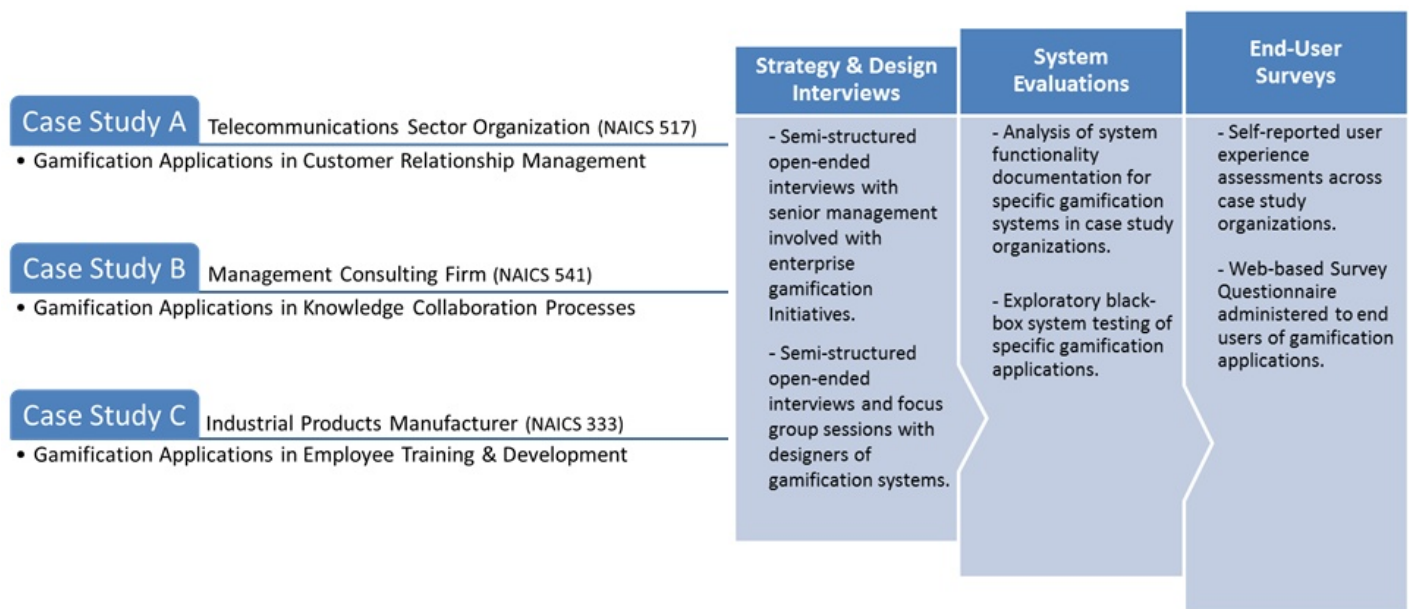

**Figure 1.** Case study organizations and data collection methods in our study

# Towards a Descriptive Framework for Meaningful Enterprise Gamification
*Umar Ruhi*

## Defining Meaningful Enterprise Gamification

As an innovative technology-based intervention, gamification entails the integration of game-like elements (game mechanics) in non-game contexts with the aim of driving positive behavioural outcomes in a target audience (Deterding et al., 2011; Hamari et al., 2014; Huotari & Hamari, 2012; Werbach, 2014). On the outset, the concept of gamification should not be confused with traditional games that are simply directed towards providing entertainment value, nor should it be mistaken for reward systems that simply entice people to perform actions to earn points. Although elements such as points, levels, leaderboards, achievements, and badges can certainly constitute components of a gamified experience (i.e., game mechanics, as described in later sections), this overall experience should be geared towards non-game situations and towards persuading end users towards intended behavioural outcomes. In the organizational context, gamification has been shown to enhance employee engagement and produce desired business outcomes in a variety of business functions including marketing, logistics, human resources, customer service, and knowledge collaboration (Buggie et al., 2014; Hense et al., 2014; Meister, 2013; Post, 2014; Sayeed & Meraj, 2013; Werbach, 2014; Wood & Reiners, 2012).

We use the term "meaningful" gamification in an enterprise context to refer to corporate scenarios where game thinking and game-based tools are used in a strategic manner to integrate with existing business processes or information systems, and these techniques are used to help drive positive employee and organizational outcomes.

Meaningful gamification should be a principal consideration for any gamification strategy to help sustain intended employee behaviours over the long term given that some early experiences of organizations have shown that, once people become bored of the gamified environments, they may not engage in the intended behaviour at all (Burke, 2014). A theoretical explanation of this phenomenon is grounded in self-determination theory (SDT) (Deci & Ryan, 2004), which suggests that, if rewards are used to encourage a behaviour that a person already has some intrinsic motivation towards, those behaviours are less likely to be observed once the rewards are removed or not perceived as valuable by that person. Hence, the key take-away for enterprise gamification is to ensure that game design elements should aim to increase intrinsic motivation among their audience. Such meaningful gamification can only be achieved with the realization that no single gamification system can cater to all users – rather, the system should be capable of providing multiple gratifications to end users, and offer features and functions that are aligned with various types of employee motivations to use the system. The next section discusses a descriptive framework that explains these factors with the aim of helping organizations think more deeply about gamification initiatives and facilitate connections between gamification application functions and end-user motivations to use those functions.

## The MDA Framework for Meaningful Enterprise Gamification

Despite the difference between traditional games and gamified systems, in defining the latter, researchers and practitioners have drawn upon formalized theoretical game design concepts such as the *mechanics, dynamics, and aesthetics* (MDA) framework (Hunicke et al., 2004; LeBlanc, 2005). *Mechanics* describe the particular rules and components of the game in terms of what actions players can undertake; the processes that drive user actions; and the conditions for progress and advancement. *Dynamics* describe how the rules manifest during actual gameplay (run-time) based on the players' inputs to the system as well as interactions among players. *Aesthetics* describe the desirable emotional responses evoked in the users when they interact with the gamified system. The MDA framework also helps in conceptualizing the relationship of the designer and the player. The designer constructs the functions and features (mechanics) of the game, which spawn different types of system–user interaction behaviour (dynamics) and lead to particular end-user emotions and experiences (aesthetics). Hence, the designer's perspective links mechanics to dynamics and subsequently aesthetics, whereas end users formulate their experiences based on the aesthetics and they engage in specific activities towards satisfying their favoured gratifications.

The MDA framework has been adopted and modified by other authors to fit the specific context of gamification, for example, the mechanics, dynamics, emotions (MDE) framework by Robson and colleagues (2015), and the design, play, experience (DPE) framework by Winn (2007). However, these and other models in the extant academic literature are primarily conceptual in nature, and to our knowledge, no empirically validated models have been published in the context of enterprise gamification. The findings from our research program aim to help address this gap.

# Towards a Descriptive Framework for Meaningful Enterprise Gamification
*Umar Ruhi*

In our preliminary research with our case study organizations, we have found the MDA framework to be a viable basis for describing the elements of enterprise gamification in a structured fashion. In our research, we have surveyed organizations from different industries utilizing gamified systems to facilitate various business practices such as customer service, knowledge collaboration, and employee training and development. Across these contexts, we have found various commonalities in the strategic requirements, system design, and user-experience elements that characterize enterprise gamification initiatives, and the MDA framework facilitates our discussion of these concepts. Our adaptation of the MDA framework is shown in Figure 2 along with empirically validated examples of mechanics, dynamics, and aesthetics that emerged in our research findings. To aid the discussion and understanding of our framework, we logically categorized the concepts in our framework as *the 20 Cs of meaningful enterprise gamification.* We do not claim that our framework comprehensively captures all aspects of enterprise gamification. It is simply an emergent framework based on specific case studies in our research program. Nonetheless, we hope that our framework offers some guiding principles for future enterprise gamification initiatives.

Additionally, our adaptation of the MDA framework incorporates the concepts of game narratives (embedded, emergent, and interpreted) that help delineate between designer and end-user perspectives of the gamification application. We explicate these concepts in the next sections by highlighting some key examples from our case studies.

Note that we deliberately use "end user" as our term of choice for consumers of enterprise gamification. Unlike in traditional games, where the term "player" is commonly used to denote a dedicated consumer role, the application consumer assumes a broader role as an employee in the context of enterprise gamification.

*Game mechanics*
At the level of game mechanics, the gamified systems we examined had very similar features and functions. *Components* such as points and badges represented basic achievements for end users who interacted with the system. For example, in the context of knowledge collaboration, a specific number of points or various types of badges would be awarded to people who have posted content or commented on questions posted by their colleagues. Leaderboards that visually display the current achievements of players in rank order were also

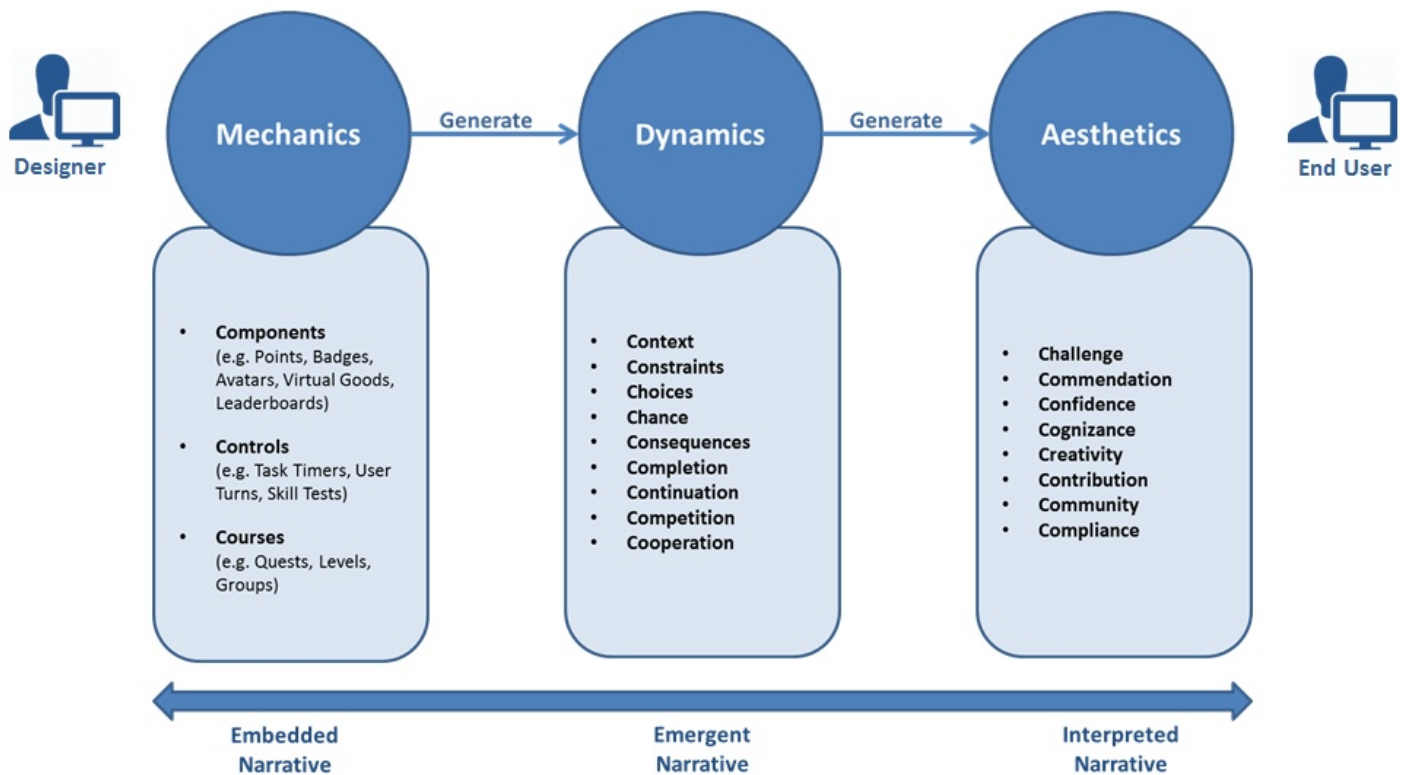

**Figure 2.** The MDA framework and the 20 Cs of meaningful enterprise gamification

# Towards a Descriptive Framework for Meaningful Enterprise Gamification
*Umar Ruhi*

fairly common among enterprise gamification systems. An example of such a system in a customer service context might entail assigning specific points for quick customer call resolutions and high customer satisfaction scores, and using these items to display the best customer service representatives on the leaderboard or to display employee dashboards with their itemized scores for various performance criteria.

Components in game mechanics are also often tied to different *courses* of action that would lead the player to higher levels on the leaderboard, and enable walkthroughs for users to allow them to unlock a sequence of relevant achievements. For instance, in training and development, completion of specific learning modules would be suggested to allow the player to proceed to the next level. Quests represent predefined challenges that typically have rewards associated with them. An example of quests that we observed in the gamified knowledge collaboration setting was the system bringing up knowledge-base articles that required further improvement or updates. These quests were linked to potential positive outcomes for the organization and often required players to collaborate with other key individuals with specific expertise in that subject area (hence incorporating group and teamwork elements).

Finally, game mechanics *controls* such as timers, turns, and tests can be used to provide cues to improve user performance. An example of controls in our study was the gamified knowledge collaboration process in which the system routinely suggested specific timelines for responding to online questions on the discussion forum, and rewarded individuals who responded within those suggested timelines. The training and development system also deployed test-based controls to facilitate employee progression across increasing stages of proficiency, and to display user accomplishments as employees overcame challenges associated with each stage.

*Game dynamics*
The game mechanics highlighted above can potentially enable different game dynamics as players interact with the gamified system. First, the *context* of the system establishes a cognitive anchoring point for players to recognize what types of activities they can undertake. For example, a monopoly-style environment for training and development that resembles the real-world board game can provide cues about specific tasks that comprise a challenge, and also encourage competition among players through a points-based system. Such a system might also have *constraints* on what players can and cannot do based on their current accumulated points and the difficulty level of the challenge. Randomness *(chance)* can also be introduced to make the gameplay more dynamic for end users, or to compel users to venture outside their comfort zones. An example of such a system that we observed in our study was an interactive customer call simulation that provided random customer complaint scenarios to be resolved through alternative means, with varied reward points associated with each step carried out by the end user. The simulation also provided dynamic feedback outlining the pros and cons of the choices made by the end users and the potential *consequences* of those choices.

The elements of *completion* and *continuation* were prevalent game dynamics across the gamification systems in our study. Progress bars indicating the proportion of completed steps in an activity or a dynamic map showing players their current and upcoming stages are some examples of such dynamics. These mechanisms help enable a sense of goal-orientation among end users and lead to feelings of satisfaction with each progress step, one notch at a time towards completion of a task or continuation to the next phase.

Together, the dynamics of consequences, completion, and continuation establish the basis for a feedback system in gamification to help drive changes in end user behaviour. Information about actions performed by end users should be linked to choices, and facilitate next steps by end users that would result in improved outcomes. As such, immediate feedback is regarded as a prerequisite to ensuring cognitive flow (i.e., a state of concentration or complete absorption with the activity at hand) (Csikszentmihalyi, 1990), which in itself is a determinant of end-user engagement.

In contrast to individual game dynamics, enterprise gamification environments also utilize collective or social dynamics including aspects of *competition* or *cooperation*. Some instances of these dynamics that we observed in our study have already been highlighted above, for example, competition among customer service representatives to achieve a higher status on the leaderboard or cooperation among subject matter experts to create or modify knowledge-base articles. Our research findings indicate that social game dynamics are more commonly exploited by end users who have relatively more experience with the gamified applications. Whether it involves working with others to achieve a mutually beneficial outcome (cooperation) or

# Towards a Descriptive Framework for Meaningful Enterprise Gamification
*Umar Ruhi*

optimizing one's own performance relative to other players (competition), social game dynamics typically require more commitment from end users and tend to operate on a longer-term basis as compared to individual game dynamics.

*Game aesthetics*
Game aesthetics represent the emotional response outcomes among end users as they participate in various activities in gamified applications. In the context of traditional games, these game aesthetics pertain to specific types of "fun" that players seek and experience during their interactions with the games, and a classification scheme for such experiences has been provided by various authors (cf. Hunicke et al., 2004; LeBlanc, 2005). In contrast to traditional games where players typically seek hedonic (entertainment or pleasure-related) gratifications, our research revealed that, in the context of enterprise gamification, end users mostly sought instrumental gratifications geared towards achieving specific valued outcomes such as learning and recognition. Hence, they saw gamification activities as a means to an end. As depicted in Figure 2, across our case studies, we uncovered eight concepts related to game aesthetics in enterprise gamification. These are briefly discussed below.

In terms of their own innate personal experiences, end users cited aspects such as challenge, confidence, cognizance, and creativity as appealing factors to participate in gamification-based activities. Many activities in gamified applications were presented in the form of *challenges* (e.g., puzzles, quizzes, difficulty levels) that required the end user to demonstrate decision-making and problem-solving skills and competencies. Through their interaction with the applications, many end users reported developing familiarity, gaining awareness, and grasping a better understanding of their business environment *(cognizance),* thinking outside the box *(creativity),* and ultimately growing their *confidence* at their workplace. A useful example of these emergent emotions and experiences was the previously highlighted simulated problematic customer call that employees needed to resolve through problem-solving skills and making dynamic decisions about next steps. End users reported that. through these exercises, they not only felt challenged to utilize their existing knowledge and skills, often in new and unanticipated ways, but the feedback provided by the gamification system also helped them understand the pros and cons of their actions and they felt better prepared to perform similar actions in their jobs.

On a more extrinsic level, end users also showed interest in gamification activities as enabling mechanisms to meet organizational standards and requirements *(compliance)* as well as to achieve recognition for their knowledge, skills, and abilities *(commendation).* For example, the completion of gamified training and development modules enabled employees to fulfil mandated training requirements, and also allowed them to showcase their credentials and be explicitly recognized for their expertise. These aspects were highly valued by end users because they often translated into immediate real-world benefits – perceived as useful "quick wins".

In addition to the self-oriented game aesthetics, our study also revealed social elements that can motivate end users to engage in enterprise gamification activities. By participating in group activities, employees reported valued emotions related to making *contributions* towards a collective goal and experiencing a sense of *community* with their colleagues in the organization. A specific instance of this in our study were employees who engaged in knowledge-collaboration activities such as answering questions on discussion forums or contributing to knowledge-base articles to document their experiences and help alleviate related problems and issues in the future. These employees reported a sense of achievement and satisfaction in helping other colleagues and their organizations.

Finally, with respect to gamification aesthetics, our analysis of end-user data across the three case study organizations rendered some key patterns in user experiences with gamification systems. As outlined in Table 1, some self-oriented aesthetics were reported with higher frequencies in the case of the gamified customer relationship management system, whereas social gratifications were more commonly reported in the case of the knowledge-collaboration system. However, emotional responses associated with confidence and cognizance were reported with high frequency across all three case study organizations. Furthermore, as highlighted earlier, some game aesthetics (especially social-oriented aesthetics) were more commonly reported by experienced end users, whereas beginners were more interested in individual game aesthetics such as commendation and compliance. Note that the relative frequencies in Table 1 are based on normalized proportions, where >60% = High; 40%–60% = Medium; and < 40% = Low. For example, in Case Study A, 24 end users were surveyed, out of which 18 cited motivations related to challenge (70%; High), 12 cited creativity (50%; Medium), and only 5 cited community (20%; Low).

# Towards a Descriptive Framework for Meaningful Enterprise Gamification
*Umar Ruhi*

Table 1. Relative frequencies (High; Medium; Low) of game aesthetics across case studies and end-user profiles

| Aesthetics in Enterprise Gamification | Case Studies | | | End-User Profiles | |
|---|---|---|---|---|---|
| | Case A: Customer Relationship Management | Case B: Knowledge Collaboration | Case C: Training and Development | Beginner End Users | Experienced End Users |
| **Challenge** | H | M | M | L | M |
| **Commendation** | H | M | H | H | H |
| **Confidence** | H | H | H | M | H |
| **Cognizance** | H | H | H | M | M |
| **Creativity** | M | M | M | L | H |
| **Contribution** | M | H | L | L | M |
| **Community** | L | H | M | L | M |
| **Compliance** | M | L | H | H | M |

The patterns in game aesthetics identified across the case studies also underline the fact that the experiences and emotional responses resulting from gamification activities are highly intertwined and not mutually exclusive. Furthermore, these experiences are highly dependent on the mindset and disposition of the players. Even within similar use cases, end users might have different referent aesthetics based on the gratifications they seek. What is common though is that end users seek these outcomes in the context of an enjoyable and fun experience, and an effective gamification platform should be able to deliver these game aesthetics within in a delightful or pleasurable manner.

## Game Narratives and Designer versus End-User Perspectives

As depicted in Figure 1, an implicit facet of the MDA framework is that it facilitates a deliberation of differences between designer and player perspectives. As shown in Figure 1, designers who create gamified applications only have direct control over the features and functions constituting the mechanics of the game, and they work with system specifications (game mechanics) that would allow specific types of user interactions (game dynamics), and ultimately meet the organizational and end-user requirements of the gamified applications (game aesthetics). On the other hand, players view the system in terms of the goals they aspire to achieve and the gratifications they receive from these enterprise gamification applications (game aesthetics). Consequently, they engage in specific gamification activities (game dynamics) drawing upon their cognitive perceptions and affective attitudes (game aesthetics) and utilize system features that offer affordances (game mechanics) to participate in their desired gamification activities.

In traditional game design, the designer and player perspectives are also often delineated in terms of narratives (Jenkins, 2003). The *embedded narrative* represents the view of the game designer in terms of structured components and event sequences intentionally embedded in a system by the designers. Hence, embedded narratives align conceptually with game mechanics. *Emergent narratives* on the other hand are created by players during their interaction with the gamification application in a dynamic fashion as they perform different activities. In this way, emergent narratives correspond conceptually to game dynamics. Finally, an *interpreted narrative* characterizes the end user's ascribed meaningfulness of experiences with the gami-

# Towards a Descriptive Framework for Meaningful Enterprise Gamification
*Umar Ruhi*

fication activities. Given that these narratives are mental representations of the players, they are logically aligned with the concept of game aesthetics.

In our research, these narratives were abundantly clear: designers and end users often spoke about the same gamification elements in different ways. For example, in the training and development gamification application, the designer inscribed the need for groups-based reward systems such as team standings and how they are different from the individual points systems (embedded narrative). On the other hand, the end users who had participated in group activities and competitions talked about aspects such as "group pride" and "team rivalries" and how these feelings allow them to perform better (interpreted narrative).

An effective gamified experience needs to be coherent across the three types of narratives, and for organizations interested in gamification initiatives, both the feature-driven perspective of the designer and the experience-driven perspective of the player are important to consider. Business requirements, user profiles, and behavioural outcomes need to be deliberated thoroughly during the planning stages of gamification initiatives, whereas technologies, tools, and tactics that would effectively engage employees in gamification activities would be key considerations during the design and implementation stages. Table 2 summarizes designer and end-user perspectives of gamification elements juxtaposed with the three game narratives.

We also analyzed game narratives at the level of game dynamics and game aesthetics with the aim of identifying patterns among these elements in terms of their most commonly reported associations (by designers and end users). Figure 3 depicts the most frequently conveyed narrative associations in the form of a bipartite graph with game dynamics and game aesthetics as its vertices. The bipartite graph is based on adjacency matrices with qualitative codes pertaining to game dynamics and game aesthetics. Edges between vertices indicate a medium or high number of co-occurrences of codes (normalized relative frequencies). The graph-

**Table 2.** Summary of gamification elements from designer and end-user perspectives

| Gamification Elements | Designer Perspective | | End-User Perspective | |
|---|---|---|---|---|
| **Mechanics** | • Objects, rules, and algorithms that need to be developed for the gamification application <br> • System specifications in terms of features and functions of the gamification platform | **Embedded Narrative** | • Gamification features and functions that act as affordances for motivational needs <br> • System features that enable performance of activities | |
| **Dynamics** | • Projected user interactions and system responses <br> • Utility of features and functions in delivering gameplay | **Emergent Narrative** | • Execution of planned activities to fulfill personal gratifications <br> • Spontaneous opportunities to participate in activities that would satisfy motivations | |
| **Aesthetics** | • Business requirements and planned user-experience outcomes from gamified systems <br> • Intended end-user responses to be evoked during gameplay | | • Motivations to engage in gamification <br> • Gratifications sought from gamified experiences <br> • Meaningfulness ascribed to gamification experiences | **Interpreted Narrative** |

# Towards a Descriptive Framework for Meaningful Enterprise Gamification
*Umar Ruhi*

based depiction offers a useful visualization aid in deciphering the prominence of different game dynamics and aesthetics.

Key highlights from this analysis include the important role of game dynamics related to *context* and *consequences*. As shown by the number of edges from these two vertices, context and consequences are key determinants in interactive gameplay, and consequently they play an important role in ensuring end-user engagement and the overall success of enterprise gamification initiatives. On the other hand, game aesthetics pertaining to *challenge* and *confidence* were reported quite frequently by end users with reference to gratifications sought from participating in gamification activities. Therefore, gamified applications need to incorporate features and cues to promote these experiential feelings among end users. Managers and designers involved in enterprise gamification initiatives should take these factors into consideration during the planning and development phases of gamification programs in their organizations.

## Guidelines for Management

Drawing upon our research findings across the three case organizations and their gamification strategies and system implementation experiences, we are able to offer the following guidelines for management of gamification initiatives.

*1. Align gamification initiatives with business objectives and intended behavioural outcomes.*
Organizations interested in enterprise gamification need to think of it as a potential method to influence specific types of behaviour in their employees. It is still early days for enterprise gamification initiatives, and the hype surrounding gamification is leading some companies to seek out ways in which they can simply use features such as points, badges, and leaderboards in a bolt-on fashion on top of existing systems. Rather, they should begin by clearly defining business objectives, formalizing planned individual and organizational outcomes, and subsequently seeking gamification solutions aligned with these objectives and outcomes.

*2. Integrate gamification strategically with business processes and information systems.*
For gamification programs to be effective, game elements should be incorporated within existing business workflows and information systems that employees use on a regular basis, and the outcomes of gamification activities should connect to desired business goals. Toward this objective, the gamification system should be used to provide feedback to employees with clear calls to action on next steps, and these systems can also help

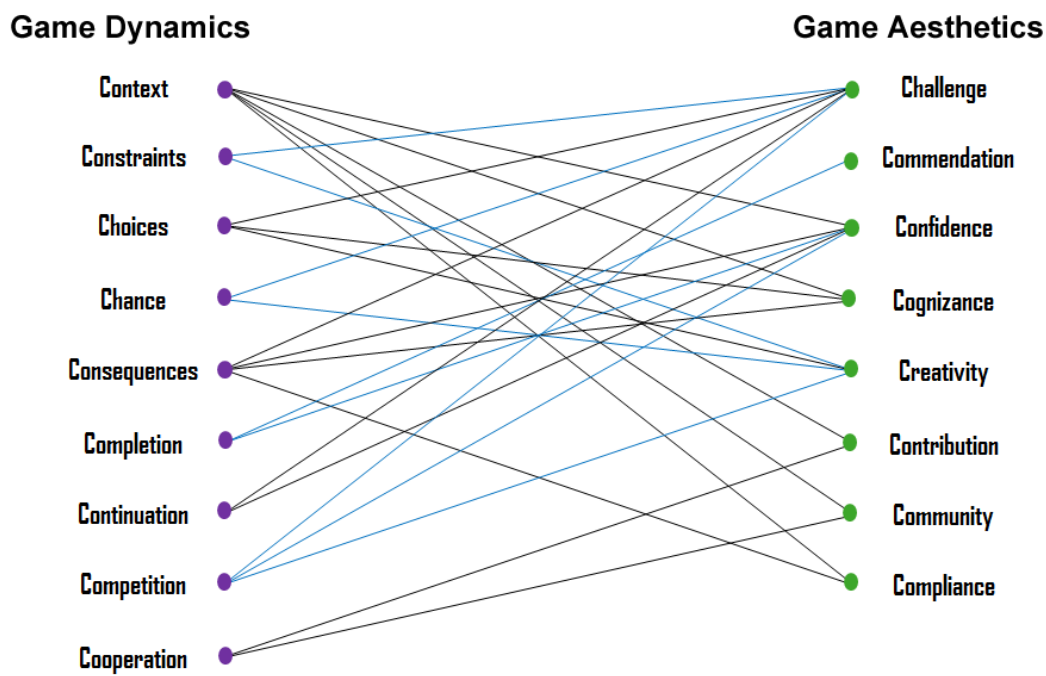

**Figure 3.** Game narratives depicted as a bipartite graph between game dynamics and game aesthetics

# Towards a Descriptive Framework for Meaningful Enterprise Gamification
*Umar Ruhi*

drive employee work compliance with corporate standards. Weaving in gamification activities and strategically placing them in the overall sequence of process events can help drive useful employee behaviours in the long term. Additionally, end-user data from gamification systems should be integrated with core information systems to allow the organization to track, reward, and recognize employees appropriately.

*3. Partner and collaborate with experts.*
Organizations need to remember that the primary purpose of building their gamification system is to engage employees and drive desired behaviours, and not to become the next great gaming company. Custom building gamified applications in-house may take more time, cost more, and entail more risk of failure as opposed to partnering with a vendor or consultant who has prior experience with building such systems and can advise on necessary requirements for success. Many vendors also provide a variety of white-label tools and customizable plug-and-play features that can help reduce the cycle time for implementation of gamification platforms.

*4. Measure and report regularly, visibly, and broadly.*
An essential component of gamification platforms is the measurement and reporting of data pertaining to end-user behaviour. Most gamification systems report such data to end users and their managers through different types of dashboards and reports. However, in order to drive long-term changes in employee behaviour, management needs to help employees understand the impact of their behaviours on the organization and visibly recognize and reward these behaviours through various offline mechanisms, perhaps using means such as corporate communication briefs or as part of employee performance reviews. Finally, metrics reported should be aligned with organizational outcomes, and it is important to communicate success often to help sustain momentum towards those outcomes.

## Guidelines for Design

Several key success factors for the design of gamification systems have already been outlined in our discussion of the MDA framework. In this section, we offer a summary of those key success factors in the form of concrete guidelines for the design of enterprise gamification applications.

*1. Design for engagement.*
At the core of the need for gamification, engagement factors into all end-user motivations and holds the key to achieving success through gamification initiatives. Designers need to ensure engagement using a variety of means such as making the gamified experience entertaining, providing stimulating challenges and rewards, and visibly linking actions and achievements to make scoring and winning transparent to the end users. Crafting a creative storified context that is linked to the work environment can help motivate individuals to participate in enterprise gamification activities. Overall, the design should provide delightful end-user experiences and results-oriented fun while enabling employees to fulfil their specific motivations.

*2. Design for personalization.*
As highlighted in our discussion on game aesthetics, end users might exhibit different and sometimes varying motivations for using gamification systems. Hence, designers need to account for these various player needs, expectations, and preferences. Toward this objective, gamified applications should offer multiple mechanisms and options to reach the same organizational objectives, and to keep people with various skill levels motivated. Additionally, applications should offer a personalized interface to end users with not just their specific game statistics, but also feedback progress reports as well as suggestions for improvement or new activities based on their profile and performance metrics. Finally, end users should be situated contextually with a relevant referent group rather than broadly in relation to the entire organization. For example, rather than using organization-wide or departmental leaderboards, gamified applications can employ segmented leaderboards according to similarities in employee profiles or based on a basket of activities that are common among a specific group of employees.

*3. Streamline the onboarding process.*
To maximize the uptake of gamification applications, their design should explicitly be geared towards minimizing barriers for end-user participation. The invitation and calls to action for playing should be clear, rules and instructions should be brief, and the interface should be simple and visually appealing. Furthermore, the first few stages of gameplay should be relatively easy and produce quick wins for end users, allowing them to assimilate the application in their routine and also to internalize an initial sense of mastery that would subsequently lead to advanced gameplay and progressive skill building.

*4. Plan, prototype, and playtest.*
Effective design begins with proper planning and cyclical improvements based on system testing and user feed-

# Towards a Descriptive Framework for Meaningful Enterprise Gamification
*Umar Ruhi*

back. Modelling using low-fidelity prototypes and storyboard mock-ups early in the design process and testing with sample end-user groups can help ensure that the gamified application would meet business objectives and satisfy individual outcomes. More formalized playtesting can be performed in later phases to allow a test group of end users to participate in gamification activities and provide their opinions. This process would be useful in identifying bugs and design flaws before releasing the application organization-wide, and would also help ensure that the application delivers the intended gameplay and spawns the desired end-user responses.

## Conclusion

Our research aims to answer the call for additional research by human–computer interaction (HCI) researchers who have stressed the need for academics and practitioners to consider features and functions of gamification technologies vis-à-vis user experience processes that drive engagement at cognitive and affective levels (Deterding et al., 2013, Nicholson, 2012). Current industry literature on this subject usually only offers advice for adding gamification as a bolt-on application or service for existing business processes (Ferrara, 2012; Zichermann & Cunningham, 2011).

Our investigation into enterprise gamification has demonstrated that an effective gamification strategy and deliberated design of gamification applications have the potential to drive key organizational initiatives. However, in order to realize the full potential of gamification and achieve effective employee engagement, organizations need to think deeply about gamification initiatives and rationalize game elements in a structured fashion rather than thinking about gamification as simply the addition of a fun videogame layer on top of existing business process systems.

The empirically validated MDA framework for enterprise gamification presented in this article may offer a viable starting point and a practical tool for organizations to conceptualize their gamification initiatives using a systematic approach. The purpose of the framework is to facilitate the selection of technology features and the design of interactive, enjoyable gameplay that would integrate well with business processes, satisfy end-user motivations, and help drive positive individual behavioural and desired business productivity outcomes – resulting in meaningful enterprise gamification.

## Recommended Reading

- *Drive* by Dan Pink offers useful background reading on the paradox of intrinsic versus extrinsic motivation. A solid understanding of these factors is a precondition for effective implementation and management of enterprise gamification initiatives. danpink.com/books/drive/

- *A Theory of Fun for Game Design* by Raph Koster describes several variations of fun that are possible in gamified systems. The book would be valuable to aspiring game designers because it helps connect the dots between game design elements and human experience outcomes. theoryoffun.com

- *The Gamification Toolkit* by Kevin Werbach and Dan Hunter offers a brief introduction to gamification by highlighting use cases and examples of game dynamics and mechanics in an enterprise setting. The book provides concise practical guidelines for managers and designers of gamified systems. wdp.wharton.upenn.edu/book/gamification-toolkit/

## About the Author

**Umar Ruhi** is an Assistant Professor of Information Systems and E-Business Technologies at the Telfer School of Management at the University of Ottawa, Canada, and a Research Associate at the IBM Centre for Business Analytics & Performance. His teaching and research interests include end-user computing, knowledge management, social computing, and consumer health informatics. His empirical research projects are predicated upon an interdisciplinary socio-technical perspective of contemporary technology applications and related organizational practices and end-user behaviour. His research projects incorporate the use of behavioural methods as well as design-science research approaches. Umar received his PhD in Information Systems from the DeGroote School of Business at McMaster University in Hamilton, Canada. His doctoral dissertation won the Best Doctoral Thesis award conferred by the German Society for Online Research (DGOF). Before joining academia on a full-time basis, Umar worked as an information technology professional and a management consultant for a variety of enterprise technology initiatives with various private and public sector organizations. More information about Umar is available at his website: umar.biz

# Towards a Descriptive Framework for Meaningful Enterprise Gamification
*Umar Ruhi*